\documentclass[twocolumn]{webofc}
\usepackage[varg]{txfonts}   
\usepackage{tikz}
\usepackage{adjustbox}
\usepackage{changepage}
\usepackage{hyperref}
\usepackage{url}
\usepackage{xcolor}
\hypersetup{colorlinks=true,citecolor=blue,urlcolor=blue,linkcolor=blue}
\begin{document}
\title{Transient size segregation of binary granular mixtures}
%
%

\author{\firstname{Soniya} \lastname{Kumawat }\inst{1}
\and
         \firstname{Anurag} \lastname{Tripathi}\inst{1}\fnsep\thanks{\email{anuragt@iitk.ac.in}} 
}

\institute{Department of Chemical Engineering, IIT Kanpur, U.P. 208016, India
          }

\abstract{Transient size segregation of a bi-disperse granular mixture flowing over a periodic chute is studied using DEM simulations and theory. 
A recently developed particle force-based size segregation model has been shown to successfully predict the steady state behavior of binary granular mixtures~\cite{tripathi2021size}. This promising model is used to predict the time-dependent segregation of different size binary mixtures in this work. 
A one dimensional continuum model is developed to solve the convection-diffusion equation by incorporating a mixture segregation model along with rheological model.
The inter-coupling of segregation with rheology is accounted to predict evolution of species concentration. 
We also investigate the effect of different initial configurations (Large near base (LNB), Small near base (SNB) and well-mixed) on the transient evolution of the flow and segregation.
The particle force-based segregation model is able to predict the evolution of the concentration profile for all three initial configurations for smallest size ratio of 1.25. \textcolor{black}{Significant deviations, however, are observable for larger size ratios, suggesting the need to account for the evolution of the velocity field in the model.} 

}
\maketitle
\section{Introduction}
\label{intro}
The segregation of different-sized grains plays a crucial role in many solid handling industries as well as geophysical situations. Segregation can be either desirable for separating particles or undesirable where mixing is required, affecting product quality. Size segregation, particularly in dense granular flows, has been extensively investigated through experimental observations~\cite{asachi2018experimental,BRANDAO20201} and Discrete Element Method simulations~\cite{chand2012discrete,combarros2014segregation,ayeni2015discrete}). 
To better understand this phenomenon, studies have focused on theoretical modelling of segregation in chutes~\cite{tripathi2021size}, bounded heaps~\cite{fan2014modelling}, plane shear flow~\cite{trewhela2024segregation}, and rotating cylinders~\cite{schlick2015granular}. Three continuum approaches for modelling segregation are commonly used in the literature: kinetic theory based approaches (e.g.~\cite{savage1988particle}), empirical methods~\cite{fan2014modelling}, and particle force-based approaches~\cite{tripathi2021size}. Among these, empirical segregation models are considered to be the most promising since they have been successful in predicting steady state as well as transient segregation due to differences in size as well as densities (see \cite{fan2014modelling,xiao2016modelling,Duan2021}). 
Very often, these empirical segregation models rely on prior knowledge of the velocity profile to determine the species concentration profiles in the mixture along with their segregation models. 
Recently, \cite{trewhela2021experimental} proposed a segregation model based on experimental observations for a single large intruder. This model is qualitatively capable of predicting the evolution of segregation behavior in various geometries, including plane shear flow~\cite{trewhela2024segregation}, chute flow, rotating square box~\cite{barker2021OpenFoam}, and triangular rotating drums~\cite{maguire2024sizetriangularrotatingOpenFoam}. These studies solve the momentum balance equations to capture the evolving flow rheology.
The particle force-based segregation model proposed by \cite{tripathi2021size} have also successfully predicted steady state segregation in periodic chutes. 
In this work, we extended the particle force-based segregation model to predict time-dependent segregation.
The continuum model predictions compare very well with DEM simulation data.

\section{Simulation Methodology}
We perform the Discrete Element Method (DEM) simulations of bi-disperse granular mixtures flowing over an inclined surface using our in-house code. We simulate sufficiently frictional (friction coefficient $\mu_{pp} = 0.5$) and slightly inelastic (coefficient of normal restitution $e = 0.88$) spherical particles of two different sizes with size ratio $r = d_L/d$ of identical densities.
We choose a simulation box with the periodic boundaries in the $x$ and $z$ directions having base area of $20d \times 20d$, where $d$ is the diameter of the small particles. Initially, the particles are arranged in a cubic lattice and allowed to settle at an inclination angle of $\theta = 0^\circ$ under the influence of gravity, until the average kinetic energy of the particles in the layer becomes below $10^{-6} mgd$. The layer height after settling is approximately $H \approx 30d$. The total number of particles in the simulation box depends on the mixture composition and ranges from $7608$ to $10536$. The inclination angle is then increased to the desired value, initiating the flow of particles. Details regarding the contact force model and property calculations can be found in \cite{tripathi2011rheology}. In order to study the effect of initial configuration on the segregation dynamics, we performed simulations for two different case of completely segregated mixtures: large near base (LNB) where all the large particles are concentrated near the base, and small near base (SNB) where all the small particles are concentrated near the base case. We also simulate a well-mixed initial configuration of particles as well. In this work, we report results for \textcolor{black}{three different} size ratios of $r = 1.25$, \textcolor{black}{$r=1.5$ and $r=2$} flowing at an inclination angle $\theta = 25^o$. 


\section{Continuum approach}
Following the previous studies~\cite{fan2014modelling}, we use the advection-diffusion equation with explicit segregation term to model segregating granular flows. The volume concentration $f_i$ of $i^{th}$ species in flowing binary granular mixture is obtained by solving the following equation using the local mixture velocity field ($\textbf{v}$).
\begin{equation}
   \frac{\partial f_i}{\partial t} +   \nabla \cdot (\textbf{v} f_i) + \nabla \cdot (\textbf{J}_{i}^{S}) + \nabla \cdot (\textbf{J}_{i}^{D}) = 0.
   \label{eq:adv_diff_seg_vectoreqn}
\end{equation}
$\textbf {J}_{i}^{S}$ and $\textbf {J}_{i}^{D}$ are segregation and diffusion fluxes of the $i^{th}$ species, respectively. For the periodic boundary condition in $x$ and $z-$ directions and no net flow in $y-$ direction, equation~\ref{eq:adv_diff_seg_vectoreqn} reduces to 
\begin{equation}
   \frac{\partial f_i}{\partial t} = -  \frac{\partial  }{\partial y} (J_{i}^{S} + J_{i}^{D}).
   \label{eq:adv_diff_seg_simplified}
\end{equation}
The segregation and diffusion fluxes are given as $J_{i}^{S} = f_i v_i^{seg}$, and $J_{i}^{D} = - D \frac{\partial f_i}{\partial y} $, $D$ being the diffusion coefficient. As observed in previous studies \cite{tripathi2021size,sahu2023particle}, we also find the linear variation of diffusivity ($D$) with the shear rate ($\dot \gamma$) along with a dependence of the local volume average diameter $d_{mix} = d [r f_L + (1 - f_L)]$. Hence, we use $D = b \dot \gamma d^2_{mix}$ in our model with $b$ as diffusivity parameter proposed in~\cite{sahu2023particle}. We use the particle force-based segregation model~\cite{tripathi2021size} to obtain the segregation velocity expression of the large species in a binary mixture as 

\begin{equation}
    v_L^{seg} = \frac{m_L g cos\theta}{c \pi \eta d_L} \alpha_0 (1-f_L)(1 + k f_L),
\end{equation}
where, $m_L$ and $d_L$ are the mass and diameter of the large particles in the mixture, respectively. $c$ is Stokes drag coefficient that seems to depend on the packing fraction~\cite{tripathi2013density}. $\alpha_0$ and $k$ are model parameters and $f_L$ is the local concentration of large particles. \textcolor{black}{We use $k = r$ (i.e., size ratio) along with $\alpha_0$ which corresponds to the ratio of net upward force to the weight of the particle, as given in \cite{tripathi2021size}.} The effective mixture viscosity ($\eta = |\tau_{yx}|/\dot \gamma$) is the ratio of shear stress ($|\tau_{yx}|$) to shear rate ($\dot \gamma$). 
By substituting the segregation velocity of large species in equation~\ref{eq:adv_diff_seg_simplified}, we obtain the segregation-diffusion equation for the large species as

\begin{equation}
   \frac{\partial f_L}{\partial t} = -  \frac{\partial  }{\partial y} \left(\frac{m_L g cos\theta \dot \gamma }{c \pi |\tau_{yx}| d_L} \alpha_0 (1-f_L)(1 + k f_L) - b \dot \gamma d^2_{mix} \frac{\partial f_L}{\partial y}  \right).
\label{eq:large_adv_diff_seg_simplified}
\end{equation}

Solving the equation~\ref{eq:large_adv_diff_seg_simplified} requires the knowledge of shear stress ($|\tau_{yx}|$) and shear rate ($\dot \gamma$). To obtain these at every time step, we need to solve time-dependent momentum balance equations. However, our DEM simulation data show that the time taken for evolution of segregation is nearly an order of magnitude larger than that for velocity. In other words, the rheological time scale is very small compared to the segregation time scale, and hence, we use the quasi-steady state momentum balance equations to obtain $\tau_{yx}$ and $\dot \gamma$ and ignore the evolution time for velocity to solve the time-dependent segregation-diffusion equation~\ref{eq:adv_diff_seg_simplified}. 

Simplified momentum balance equations for steady fully developed flow over an inclined plane at an inclination angle $\theta$ in $x$ and $y-$ directions are given as 
\begin{equation}
    \tau_{yx}(y) = - \phi \rho_p g \sin \theta  (H - y),
    \label{eq:mombal_tau}
\end{equation}
\begin{equation}
    P (y) =\phi \rho_p  g \cos \theta (H - y),
    \label{eq:mombal_p}
\end{equation}
where $\rho_p$ is the particle density, $\phi$ denotes local packing fraction, which is assumed to remain constant both in time and space, with a value of $\phi = 0.58$. Dense flow rheology dictates that the pressure and shear stress relate to each other through an inertial number dependent effective friction coefficient, expressed as $\mu(I_{mix}) = \frac{|\tau_{yx}|}{P}$ which equals to $\tan \theta$ using equations~\ref{eq:mombal_tau} and \ref{eq:mombal_p}. Further, we compute the inertial number by rearranging the empirical form of $\mu(I)$ given by \cite{jop2006constitutivenature} as
\begin{equation}
I_{mix} =I_{o} \dfrac{\mu(I_{mix})-\mu_{s}}{\mu_{m}-\mu(I_{mix})}.
\label{eq:Imix_mui}
\end{equation}
Here, $I_{o}, \mu_{s}, \mu_{m} $ are the rheological parameters, and the values for these parameters are same as given by \cite{tripathi2011rheology}. Next, we calculate the shear rate $\dot{\gamma}$ using the dimensionless inertial number $I_{mix}$ expression, $\dot{\gamma}\textcolor{black}{(y)}=I_{mix}d_{mix} / \sqrt{P / \rho_{_{P}}}$.
After substituting $\dot \gamma$ and $|\tau_{yx}|$ (using equation~\ref{eq:mombal_tau}), we solve equation~\ref{eq:large_adv_diff_seg_simplified} to obtain the concentration of large species using the MATLAB pdepe solver. The required initial and boundary conditions are given as follows:
\begin{equation}
\begin{split}
     f_{L}(y,0) = f_{L,ini}(y),\\
     J_{L}^{S}(0,t) + J_{L}^{D}(0,t) = 0,\\
     J_{L}^{S}(h,t) + J_{L}^{D}(h,t) = 0.
    \end{split}
\label{eq:IC_BC_concentration}
\end{equation}
Here, the initial concentration profile of the $i^{th}$ species, $f_{L,ini}(y)$, is equal to the total composition of that species in the mixture ($f^T_L$) in case of well-mixed initial configuration. For a near complete segregated mixture, $f_{L,ini}(y)$ is represented by a step function. 

\begin{figure}[h]
    \centering
    \includegraphics[scale=0.25, trim=0 0 0 0, clip]{Figures/concen-r-1.25-fl-50.eps}\put(-120,90){(a)}
    \hspace{-0.55cm} 
    \begin{tikzpicture}
        \node[anchor=south west] (base1) at (-11,0) 
        {\includegraphics[scale=0.25, trim=0 0 0 0, clip]{Figures/com-r-1.25-fl-50.eps}
            };
        \node at (-11,3.2) {(b)};
        \node[anchor=south west] at (-9.9,2.2) 
   {\includegraphics[scale=0.04, trim=200 0 30 0, clip]{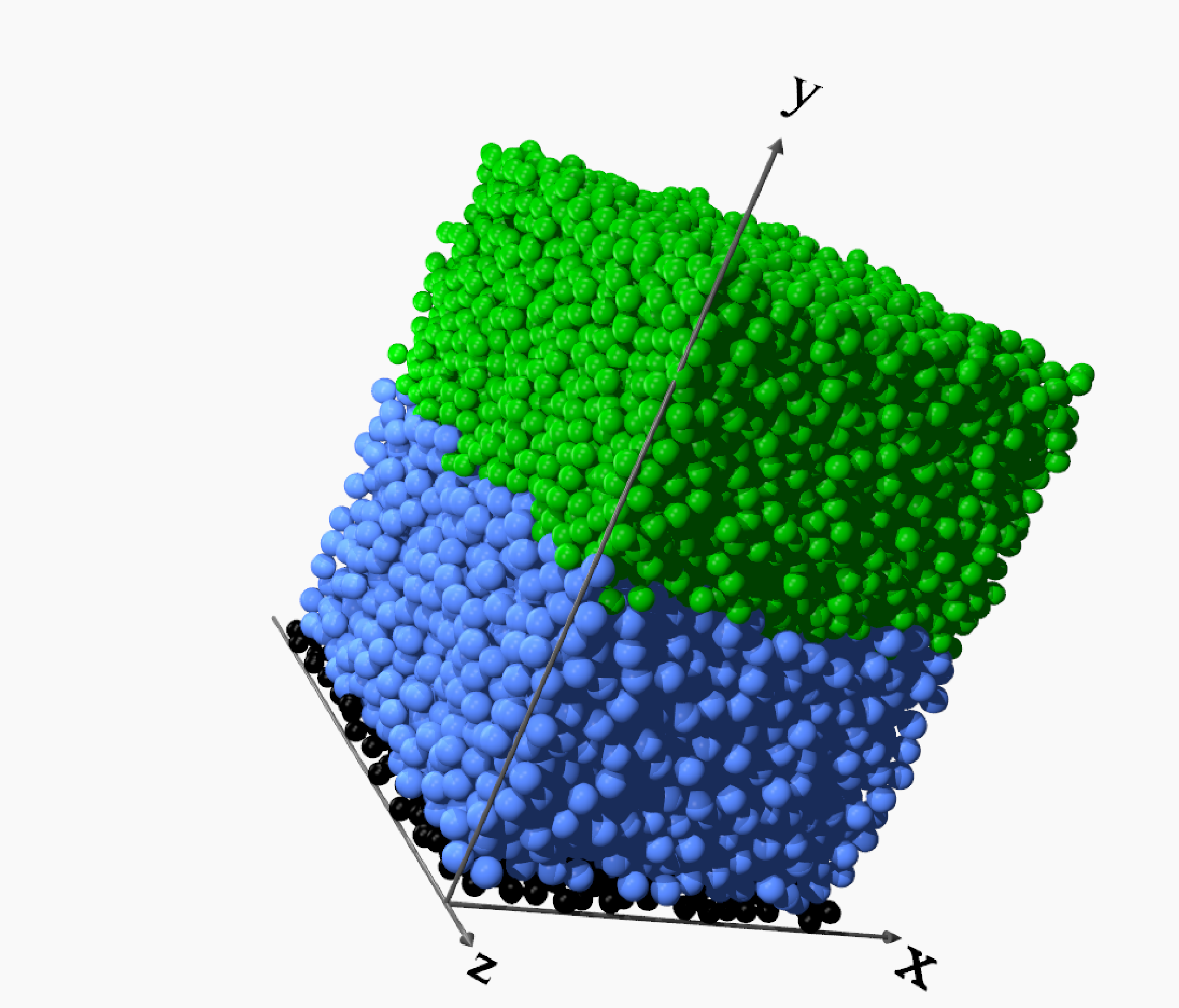}};
    \end{tikzpicture}
    \caption{(a) Instantaneous concentration profiles of large species in an equal composition binary mixture having size ratio ($r  = 1.25$) flowing at inclination angle $\theta = 25^o$. (b) Time variation of centre of masses ($y_{com}$) for both large (blue) and small (green) species. 
    }
\label{fig:r_1.25_theta_25_fl_50instant}
\end{figure}

\section{Results and discussion}
\label{sec-1}
In this section, we report the comparison of continuum model predictions of the transient evolution of species concentration with the DEM simulation data. \textcolor{black}{The results are reported in non-dimensional form, where both $y$ and $y_{com}$ are scaled by the small particle diameter $d$ and time $t$ is scaled by $\sqrt{d/g}$.} Figure~\ref{fig:r_1.25_theta_25_fl_50instant} shows the results for $50\% - 50\%$ binary mixture (size ratio $r = 1.25$) flowing at an inclination angle of $\theta = 25^o$. The flow begins from a large near base (LNB) configuration (snapshot shown in figure~\ref{fig:r_1.25_theta_25_fl_50instant}b -inset). 
Figure~\ref{fig:r_1.25_theta_25_fl_50instant}a shows the instantaneous concentration profiles of large species at different times. Initially, the large particles are concentrated near the base (LNB case), and as the flow evolves, they gradually move toward the free surface, while the smaller particles move towards the base. Large particles are observed in the top layer at the steady state with clear dominance of small particles near the bottom. \textcolor{black}{The model predictions (lines) match with the corresponding DEM data (symbols of the same color) except near the free surface and the base due to the lower values of the solids fraction in these regions.}
The centre of mass ($y_{com}$) of large and small species for this case is shown in figure~\ref{fig:r_1.25_theta_25_fl_50instant}b. 
$y_{com}$ of the large particles (blue) increases over time, while that of the small particles (green) decreases. Solid lines (model predictions) closely match symbols (DEM data) indicating that the continuum model predicts the temporal evolution of segregation dynamics \textcolor{black}{reasonably well}.

\begin{figure}[h]
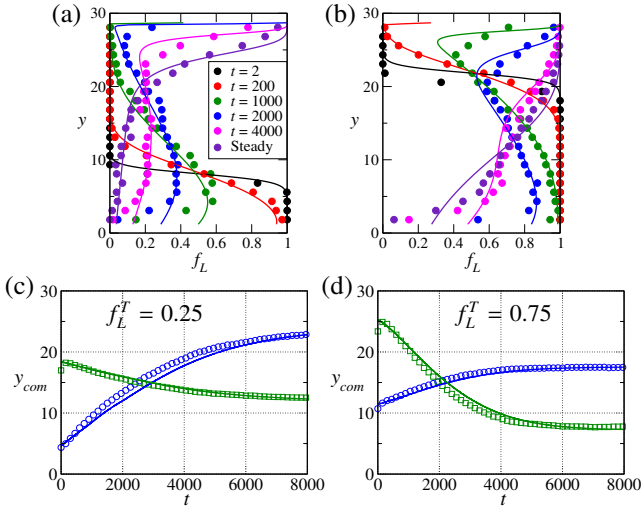

    \centering
     \includegraphics[scale=0.27, trim=0 0 0 0, clip]{Figures/concen-r-1.25-fl-25.eps}\put(-90,95) {(a)} ~~~~~~~
    \includegraphics[scale=0.27, trim=0 0 0 0, clip]{Figures/concen-r-1.25-fl-75.eps}\put(-90,95) {(b)}
    \hfill
    \includegraphics[scale=0.25, trim=0 0 0 0, clip]{Figures/com-r-1.25-fl-25.eps}\put(-120,80){(c)}\put(-82,70){\small $f^T_L = 0.25$}
    \includegraphics[scale=0.25, trim=0 0 0 0, clip]{Figures/com-r-1.25-fl-75.eps}\put(-120,80){(d)}\put(-70,70){\small $f^T_L = 0.75$}
    \caption{Instantaneous concentration profiles of large species ($f_L$) in binary mixture with size ratio $r  = 1.25$ for two different  mixture compositions: (a) $f^T_L = 0.25$, and (b) $f^T_L = 0.75$. The corresponding time variation of $y_{com}$ shown in (c), and (d) respectively. 
    }
\label{fig:r_1.25_theta_25_flt25_75}
\end{figure}

Next, we report the effect of mixture composition on the segregation dynamics of binary size mixtures. Figure~\ref{fig:r_1.25_theta_25_flt25_75} shows the results for two different mixture compositions for size ratio $r = 1.25$ at inclination $\theta = 25^\circ$. As in figure~\ref{fig:r_1.25_theta_25_fl_50instant}a, flow starts from large near base configuration. 
Figure~\ref{fig:r_1.25_theta_25_flt25_75}a shows the instantaneous concentration profiles of large species for a mixture rich in small particles ($f^T_L = 0.25$) at different times. Similarly, figure~\ref{fig:r_1.25_theta_25_flt25_75}b shows the instantaneous concentration profiles of large species in mixture rich in large particles ($f^T_L = 0.75$). The corresponding $y_{com}$ data are shown in figures~\ref{fig:r_1.25_theta_25_flt25_75}c and \ref{fig:r_1.25_theta_25_flt25_75}d.
The continuum model predictions match the DEM simulation data accurately for both low and high compositions of large particles in the mixture.

\begin{figure}[h]
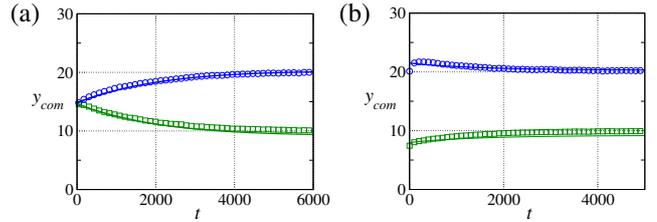

    \centering
    \includegraphics[scale=0.24, trim=0 0 0 0, clip]{Figures/Mixed-com-r-1.25-fl-50.eps}\put(-120,75) {(a)} \hfill
    \includegraphics[scale=0.24, trim=0 0 0 0, clip]{Figures/SNB-com-r-1.25-fl-50.eps}\put(-115,75) {(b)} \hfill
    \caption{Time variation of $y_{com}$ for $50\% - 50\%$ mixture of size ratio $r = 1.25$ starting to flow from (a) well-mixed case, and (b) small near base. 
    }
    \label{fig:diff_Initial_conf}
\end{figure}

Next, we validate the continuum model by comparing its predictions with the DEM data for two distinct initial configurations.
The mixture considered is a $50\% - 50\%$ binary mixture flowing at an inclination angle of $\theta = 25^\circ$ and we only report the time variation of species centre of mass ($y_{com}$). 
Figure~\ref{fig:diff_Initial_conf}a shows the evolution of $y_{com}$ starting from a position of well-mixed case. The initial values of $y_{com}$ are identical for both species at $t = 0$. With time, the position of center of mass ($y_{com}$) of the large (blue) particles increases, while that of the small (green) particles decreases. Similarly, figure~\ref{fig:diff_Initial_conf}b shows the evolution of $y_{com}$ for the initially segregated state with large particles occupying at the top half. In this case, only diffusive mixing occurs at the interface, causing a slight change in $y_{com}$ with time.

The steady state values of $y_{com}$ for both large and small particles in figures~\ref{fig:diff_Initial_conf}a and \ref{fig:diff_Initial_conf}b are the same as those observed in the LNB case (shown in figure~\ref{fig:r_1.25_theta_25_fl_50instant}b).
Thus, although the extent of segregation at steady state remains unchanged for initial configurations, the time required to reach steady state and the evolution of segregation are significantly influenced by the initial configuration. Again, the continuum model accurately captures the segregation dynamics, with its predictions (solid lines) closely matching the DEM data (symbols) for all different initial configurations.

\begin{figure}[h]
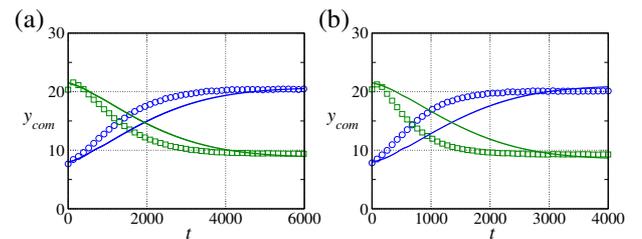

    \centering
    \includegraphics[scale=0.24, trim=0 0 0 0, clip]{Figures/com-r-1.5-fl-50-1eqnmodel.eps}\put(-115,80) {(a)} 
   \includegraphics[scale=0.24, trim=0 0 0 0, clip]{Figures/com-r-2.0-fl-50-1eqnmodel.eps}\put(-115,80) {(b)} 
    \caption{Time variation of $y_{com}$ for $50\% - 50\%$ mixture for size ratios (a) $r = 1.5$, and (b) $r = 2.0$ starting from large near base initial configuration.  
    }
    \label{fig:r_1.5_2.0}
\end{figure}

\textcolor{black}{We next report the results for larger size ratios  $r = 1.5$ and $2.0$, shown in figures~\ref{fig:r_1.5_2.0}a and \ref{fig:r_1.5_2.0}b, respectively. The continuum model does not capture the evolution of $y_{com}$ for these larger size ratios. This is due to the fact that the continuum model assumes that the rheological time scale is much shorter than the segregation time scale. Increase in size ratio leads to faster segregation (and hence decreased segregation time scale), making the assumption less accurate for larger size ratios. Since the steady state segregation is accurately predicted using this model, we conclude that the accounting of the velocity evolution with time is critical for these size ratios. }
\section{Summary}
\label{sec_summary}
We study the size segregation of dense binary granular mixtures of spherical particles flowing over an inclined plane and developed a continuum model that predicts the segregation dynamics. We utilize the particle force-based segregation model proposed by \cite{tripathi2021size} and incorporate the steady state velocity field through the $\mu(I)$ rheological model~\cite{jop2006constitutivenature}.  Our DEM simulation data reveals that the rheological time scale is significantly smaller than the segregation time for the small size ratios. 
\textcolor{black}{This simplified model predicts the segregation dynamics in good agreement with DEM simulation data across the bulk region of layer over a wide range of mixture composition for three different initial configurations considered in this study. 
The model is unable to capture the segregation in the basal region and near the free surface due to lower values of the solids fraction in these regions.}
\textcolor{black}{However, the continuum model predictions differ significantly from the DEM results for larger size ratios of $1.5$ and $2$. This assumption is more accurate for small size ratios and becomes less accurate for larger size ratios. The accounting of the velocity evolution seems to be important for these size ratios and will be explored in future.
}
\section*{Acknowledgments}
AT and SK gratefully acknowledge the funding support provided to SK by the Prime Minister's Research Fellowship (Government of India) grant.

\end{document}